\documentclass[11pt]{article}

\usepackage{geometry}
\usepackage{currfile}
\usepackage{textcomp}
\usepackage{enumitem}

\usepackage{url}
\usepackage{doi}

\usepackage{xcolor,soul}
\sethlcolor{yellow}

\usepackage[pdftex]{graphicx}
\DeclareGraphicsExtensions{.pdf,.jpeg,.png}
\graphicspath{{./figures/}}

\date{October 7, 2023}

\begin{document}

\title{\Large{Data Cooperatives for Identity Attestations  \\
~~\\}}
\author{
\large{Thomas~Hardjono~~~~Alex Pentland}\\
\large{~~}\\
\large{MIT Human Dynamics \& Connection Science}\\
\large{~~}\\
\small{{\tt hardjono@mit.edu}~~~~{\tt pentland@mit.edu}}\\
\large{~~}\\
}

\maketitle

\begin{abstract}
Data cooperatives with fiduciary obligations to members provide
a useful source of truthful information regarding a given member
whose personal data is managed by the cooperative.
Since one of the main propositions the cooperative model
is to protect the data privacy of members,
we explore the notion of {\em blinded attestations}
in which the identity of the subject is removed 
from the attestations issued by the cooperative regarding
one of its members.
This is performed at the request of the individual member.
We propose the use of a legal entity to countersign the blinded attestation,
one that has an attorney-client relationship with the cooperative,
and which can henceforth become the legal point of contact
for inquiries regarding the individual related to the attribute being attested.
There are several use-cases for this feature,
including the Funds Travel Rule in transactions in digital assets,
and the protection of privacy in decentralized social networks.
~~\\
\end{abstract}

\section{Introduction}
\label{sec:Introduction}

As we described in~\cite{PentlandHardjono2020-DataCooperativesChapter}
the notion of a {\em Data Cooperative} refers to the voluntary collaborative
pooling by individuals of their personal data for the benefit
of the members of the group or community. 
A key motivation for 
individuals to get together and pool their data is driven by the
need to share common insights across data that would otherwise
be siloed or inaccessible. 
These insights provide the members with a better understanding of their current
economic, health, and social conditions as compared to the other
members of the cooperative in general.

Given the low-cost and high available of IT technologies today (e.g. cloud services,
Software-as-a-Service (SaaS) platforms, etc),
it is technically straightforward to have a third party such as a
cooperative hold copies of their members’ data.
This is in order to help them safeguard their rights, represent them in negotiating how
their data is used, alert them to how they are being surveilled,
and audit the large companies and government institutions
using their members’ data. 
The creation of such data cooperatives also does not necessarily require new laws.
Many community organizations are already chartered
to manage members’ personal information for them. 
It does,
however, require new regulations and oversight, similar to how
the government regulates and provides oversight of financial institutions.

There are several key aspects to the notion of the 
data cooperative~\cite{HardjonoPentland2019-DataCooperatives-Arxiv}:
\begin{itemize}

\item	{\em Individual members own and control their personal data}:
The individual as a {\em member} of the data cooperative
has unambiguous legal ownership of (the copies of) their data.
This data is added into the member's
{\em personal data store} (PDS)~\cite{openPDS2014PLOS}.

\item	{\em Fiduciary obligations to members}:
The data cooperative has a legal fiduciary obligation 
first and foremost to its members~\cite{Balkin2016}.
The organization is member-owned and member-run,
and it must be governed by rules (bylaws)
agreed to by all the members.

\item	{\em Improve the lives of members}:
The goal of the data cooperative is to benefit its members 
first and foremost.
The goal is not to ``monetize'' their data, 
but instead to improve the members' lives through a better
understanding of their current economic, health, and social conditions.

\end{itemize}


\section{Motivations: Improving Access to Identity Attributes}
\label{sec:Motivations}

A major benefit for individuals
in participating their personal data to the private pool of data managed 
by the cooperative pertains to the ability
for the cooperative to carry-out algorithmic computations 
that yield insights about the individual and about the community as a whole
regarding some aspects of their lives
(e.g. health, financial, etc.).
Regulations exist today in several industries (e.g. health, finance, telecoms, etc.)
that permit individuals to request copies of their data legally
(e.g. medical history files, credit card and bank transactions data,
telco call-data-records and location data, etc).
Most individuals do not have the technical skills to store and manage these data sets,
let alone perform algorithmic computations on them.
Hence the need for the data cooperative model
where members trust the cooperative,
which operate under a fiduciary relationship with its members.

Extending from this fiduciary relationship is the ability of the cooperative
to validate specific attributes of a member
(e.g. age, location of residence, income bracket, etc.) under the directive and consent of the member,
and to attest to these attributes.
An individual member can, therefore, request the cooperative 
to issue signed attestations in a digital format that can later
be utilized by the individual to obtain services at third parties
who need to perform risk assessment about that individual
(e.g. car loan from a bank or credit union).

The data cooperative model is especially attractive today given
the backdrop of poor data privacy practices around the world:
\begin{itemize}

\item	{\em Declining trust in institutions}: 
Over the last decade there has been a continuing decline in trust on the part of individuals
with regards to the handling and fair use of personal data.
Pew Research reported that 91 percent
of Americans agree or strongly agree that consumers 
have lost control over how personal data is collected and used, 
while 80 percent of those who use social networking sites are 
concerned about third parties accessing their shared data~\cite{Madden2014}. 
The Webbmedia Group, writing in the Harvard Business Review, 
has identified data privacy as one of the top ten technology trends this decade~\cite{Webb2015,Maler2015a}.
This situation has also been compounded by the various 
recent reports of attacks and theft of data
that directly impacted the citizen
(e.g. Anthem~\cite{Anthem2015}, Equifax~\cite{Equifax2017}).

\item	{\em Privacy is inadequately addressed}: 
The 2011 WEF report on personal data as a 
new asset class~\cite{WEF2011} finds that the current ecosystems that access and use personal 
data is fragmented and inefficient.  For many participants, the risks and 
liabilities exceed the economic returns and personal privacy concerns are 
inadequately addressed. Current technologies and laws fall short of providing 
the legal and technical infrastructure needed to support a well-functioning 
digital economy. The rapid rate of technological change and commercialization 
in using personal data is undermining end-user confidence and trust.

\item	{\em Regulatory and compliance requirements}: 
The introduction of the EU General 
Data Protection Regulations (GDPR)~\cite{GDPR,OAIC-Australia-2018} 
will impact global organizations that rely 
on the Internet for trans-border flow of raw data. This includes cloud-based 
processing sites that are spread across the globe.

\end{itemize}
More recently,
the approval of the EU~MiCA Regulation~\cite{MiCA2023} on digital assets
highlights the degree to which the Web3 digital assets industry
is ill-prepared to address the regulatory demands
regarding entity identity information.

\begin{figure}[t]
\centering
\includegraphics[width=0.9\textwidth, trim={0.0cm 0.0cm 0.0cm 0.0cm}, clip]{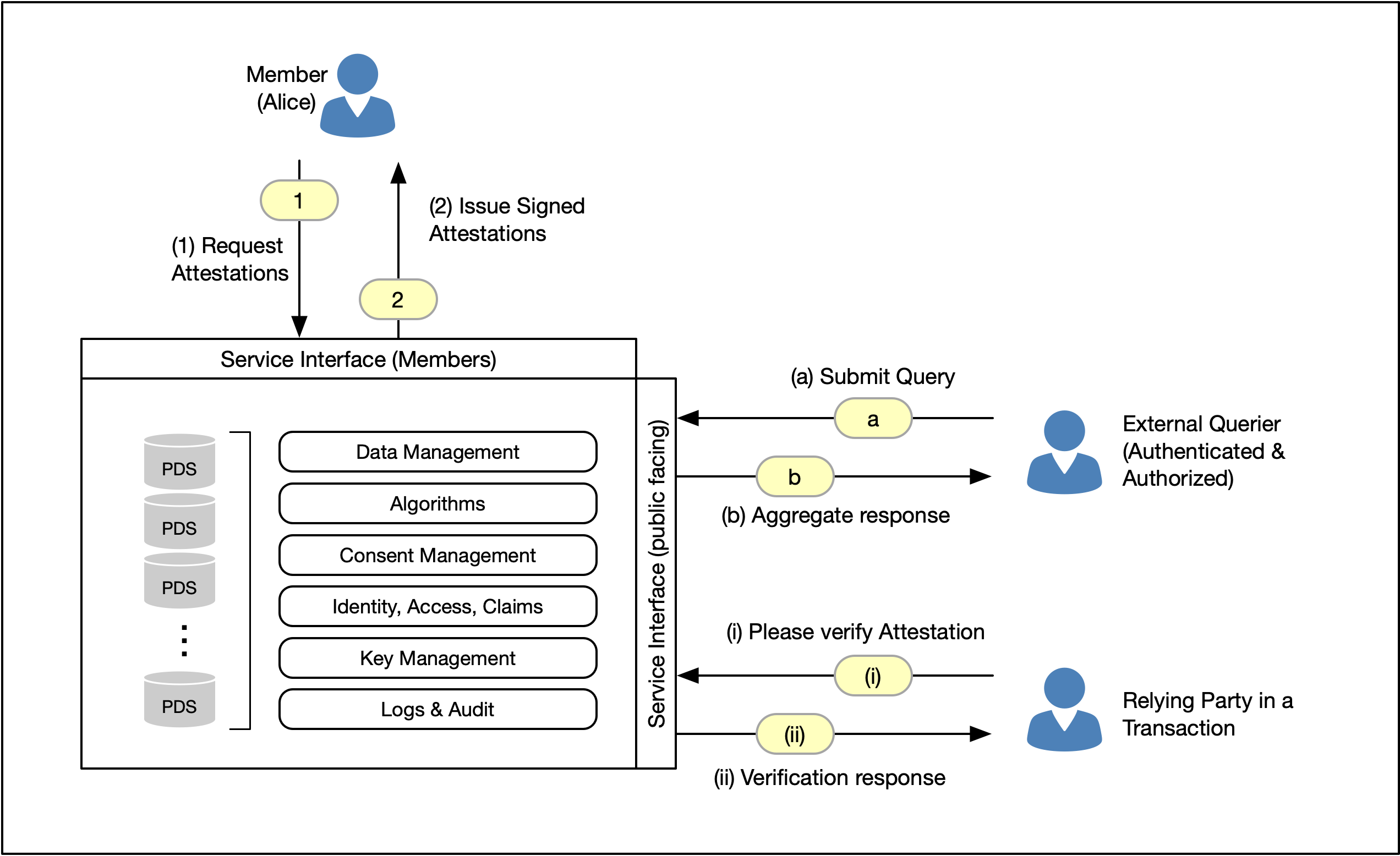}
\caption{The Data Cooperative Model}
\label{fig:datacoop-model}
\end{figure}


\section{The Data Cooperative as Source of Identity Attestations}
\label{sec:CoopAttester}

In the past few years, the notion of a ``self-sovereign identity'' has emerged in the context of discussions
around digital identity, data privacy, and control over an individual's personal data. 
Although it has a catchy sound, the term may be misleading because, among others,
it does not reflect the reality that the majority of individuals live and interact
within real-world (physical) communities~\cite{Pentland2015-short}.
As such,
the term {\em community sovereign} may be a more 
accurate description of the situation~\cite{PentlandHardjono2018a}.

Given this reality the community itself is perhaps a better source
of information regarding a person and their reputation within the community. 
The traditional paper Letters of Introductions from the community
were commonly utilized (e.g. for traveling) for the most part
of several millennia, 
prior to the arrival of digital technology.
Analogously, a ``digital letter of introduction'' issued by 
a data cooperative representing a community 
should be a reliable source of {\em attestations} regarding
an individual member of that community, 
disclosing attributes about the individual and other related reputation information.

In the digital space,
one of the earliest attempts to digitally codify the relationship 
between an individual and their community
was the {\em Pretty Good Privacy} (PGP) project in 1991
in the context of public-keys~\cite{RFC1991-formatted}.
The basic idea is the same: 
members of a community would vouch for an individual's name (as a member of that community)
and possession of a public-key
by way of counter-signing the individual's public-key.
They would then add this public-key to their respective ``PGP key-rings''.
The so-called ``PGP key signing ceremonies'' would also be conducted in public,
usually at technical events such as the IETF meetings that occur three times per year
since the 1990s
(e.g. see~\cite{PGP-Key-Signing-IETF-1999}).
Such a ceremony (usually after dinner) would be attended and witnessed by
the members of the specific PGP community,
and in many cases also by the broader attendees of the event
(who may not even be familiar with the individual in question).

We therefore see the data cooperative as an instance of community-sovereign formation,
where human relationships are expressed not merely in the digital space,
but instead in daily interactions in the real world
among the members of that community.
This real-world engagement among people is akin to the PGP key signing ceremonies
which are attended by real people in the physical world.

Figure~\ref{fig:datacoop-model} provides a high-level illustration
of the data cooperative model,
with a member (Alice) requesting the data cooperative
to issue and sign attestations regarding that member (Step-1 and Step-2).
The computed attestations can be expressed in various syntaxes,
including the {SAML2.0}~Assertions format~\cite{SAMLcore},
the W3C Verified Credentials format~\cite{Sporny2022},
the OpenID-Connect ID-token format~\cite{OIDC1.0},
and others.

In Figure~\ref{fig:datacoop-model} once Alice delivers
the issued attestations to the intended recipient 
(referred to as the {\em Relying Party} (RP)~\cite{SAMLglossary,ABA2012}),
the RP can either accept the signature of the cooperative
or request the cooperative to re-validate the statement
in the signed attestations (Step-(i) and Step-(ii)).
Since each issued attestation has an expiration date,
this re-validation may be needed to ensure that the attestation
has not been revoked by the cooperative
before its stated expiration date.

\newpage


\section{Blinded Attestation and Attorney-Client Privilege}
\label{sec:Blinding}

There are several emerging Web3 scenarios that require attributes regarding
a person (i.e. data subject~\cite{GDPR}) to be attested 
without immediately disclosing the identity of the subject.
That is, the subject remains anonymous until such time
their identity must be disclosed in order for the transaction to proceed to the next stage.

An example of this need is related to the trading of digital assets
on decentralized networks (e.g. blockchain-based, DLTs)
where the identity of the Originator and the Beneficiary
must be known in order for the transaction
to comply to the existing financial regulations 
(e.g. AML/KYC~\cite{FATF-Recommendation15-2018}).
In the case of public/permissionless blockchains,
the direct publishing of person-identifying data (e.g. name, address, phone, etc)
on the blockchain may negatively affect 
not only the data privacy of the Originator and Beneficiary (individuals or organizations)
but may also affect their physical safety (e.g. from criminal threats).

The need for user anonymity is often
dictated by the various aspects -- often seemingly contradictory --
related to the context of the transaction.
These aspects include personal data privacy,
regulatory enforcement in certain jurisdictions,
preventing identity theft, and others.
More recently,
the topic of anonymity (pseudonymity) of users has returned to the foreground most
notably in the context of cryptocurrencies (e.g. Bitcoin).
However,
the interest in digital identity anonymity pre-dates blockchains and cryptocurrencies,
and researchers have been exploring various identity anonymity schemes
for the past three decades (e.g. see~\cite{Chaum81,Brands93a,Camenisch2002,BrickellLi2012}).
In many circumstances, a trade-off must be made between
the need to conduct day-to-day transactions (e.g. payments)
versus the practicality of many of these cryptographic identity anonymity schemes.
Most (all) of these identity anonymity schemes
have not undergone the rigorous technical standardization process
and have yet to experience extensive field deployment (which typically yields bugs and design errors).
The same applies for recent proposals based on cryptographic zero-knowledge proof (ZKP) schemes.

\begin{figure}[t]
\centering
\includegraphics[width=0.9\textwidth, trim={0.0cm 0.0cm 0.0cm 0.0cm}, clip]{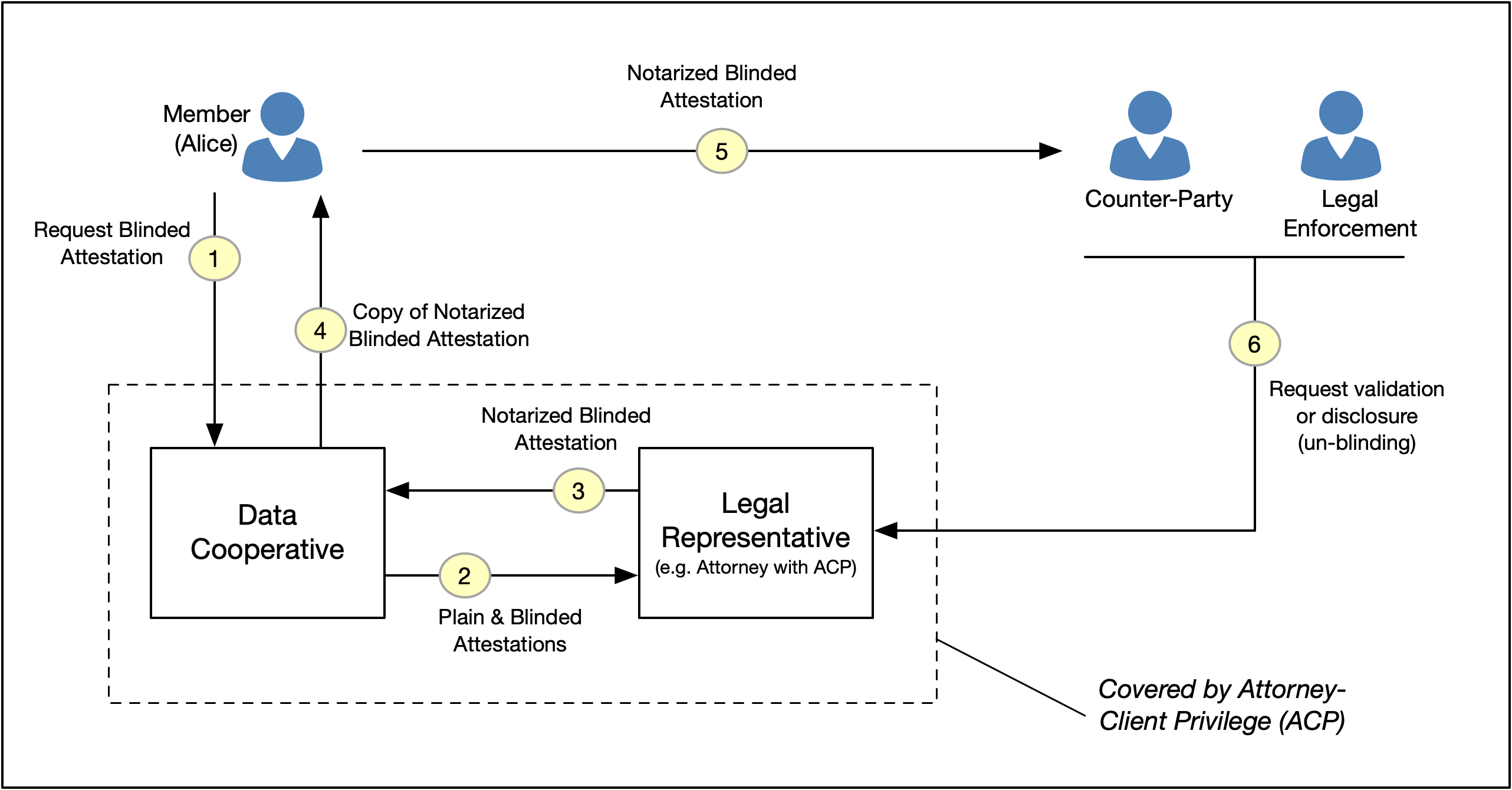}
\caption{Overview of the blinded attestations with attorney-client privilege (ACP)}
\label{fig:blinding}
\end{figure}

In many ordinary deployment scenarios what is often required
is not unconditional anonymity 
(e.g. with true {\em untraceability} and {\em unlinkability}~\cite{Camenisch2002})
but rather a temporary hold on identity-attribute disclosure pending
a legal demand for such disclosure.
In other words,
the goal of the user is not to achieve perfect anonymity
but to disclose their identity only to relevant parties in the transaction
upon request (e.g. from legal enforcement bodies).
We propose the use of a simple and pragmatic legal {\em attestation blinding}
which can be achieved with the involvement of a legal representative of 
the data cooperative that is covered under {\em attorney-client privilege}~\cite{ABA1983-Attroney-Client}.
This is summarized in Figure~\ref{fig:blinding}.

Our notion of attestation blinding\footnote{We borrow the term ``blinding'' 
from the classic work of Chaum on blinded electronic cash~\cite{Chaum:1990:UEC:88314.88969}.}
is where a legal representative of the cooperative (or of the individual member)
performs the simple blinding process,
acting also as a {\em digital notary}.
The legal representative (e.g. Law Firm) must operate within a jurisdiction that
recognizes attorney-client privilege (ACP) with the client,
which in this case would be the cooperative and its members. 
As we discuss later in Section~\ref{sec:FATF},
this attorney-client relationship is beneficial to the subject
in the context of digital asset transactions in decentralized asset networks
(e.g. public blockchains).

This attestation blinding workflow is summarized
in Figure~\ref{fig:blinding}.
In Step-1 the member (Alice) requests the cooperative
to issue a blinded attestation in which only the attributes
of the member are shown or listed (no member identification or other identifying information).
The cooperative computes two (2) versions of the attestation.
The first is the plain (unblinded) attestation.
The second is the blinded attestation,
which carries a cryptographic hash of the plain attestation.
Both are digitally signed by the cooperative,
and both are then delivered to the legal representative under ACP coverage.
This is shown as Step-2 in Figure~\ref{fig:blinding}.

On its part,
the legal representative compares the plain attestation and the blinded attestation
to ensure that they match (i.e. the same attribute being presented,
the hash is valid, the timestamp is valid, etc.).
It then countersigns (append its signature) to the blinded attestation
before returning it to the cooperative (Step-3).
It also logs and archives a copy of all three data structures
(the plain attestation, the blinded attestation and 
the countersigned blinded attestation).
The current industry standard for digital signatures
supports enveloping and countersigning
(e.g. for signature standards see
see~\cite{NIST-DSS-2023,RFC5652-CMS-Formatted,RFC7517-JWK-Formatted,RFC7515-JOSE-Formatted}). 
These technical standards for digital signatures and enveloping
have been deployed in the computer and network
industry for over two decades,
and therefore well-understood and broadly deployed.

\begin{figure}[t]
\centering
\includegraphics[width=1.0\textwidth, trim={0.0cm 0.0cm 0.0cm 0.0cm}, clip]{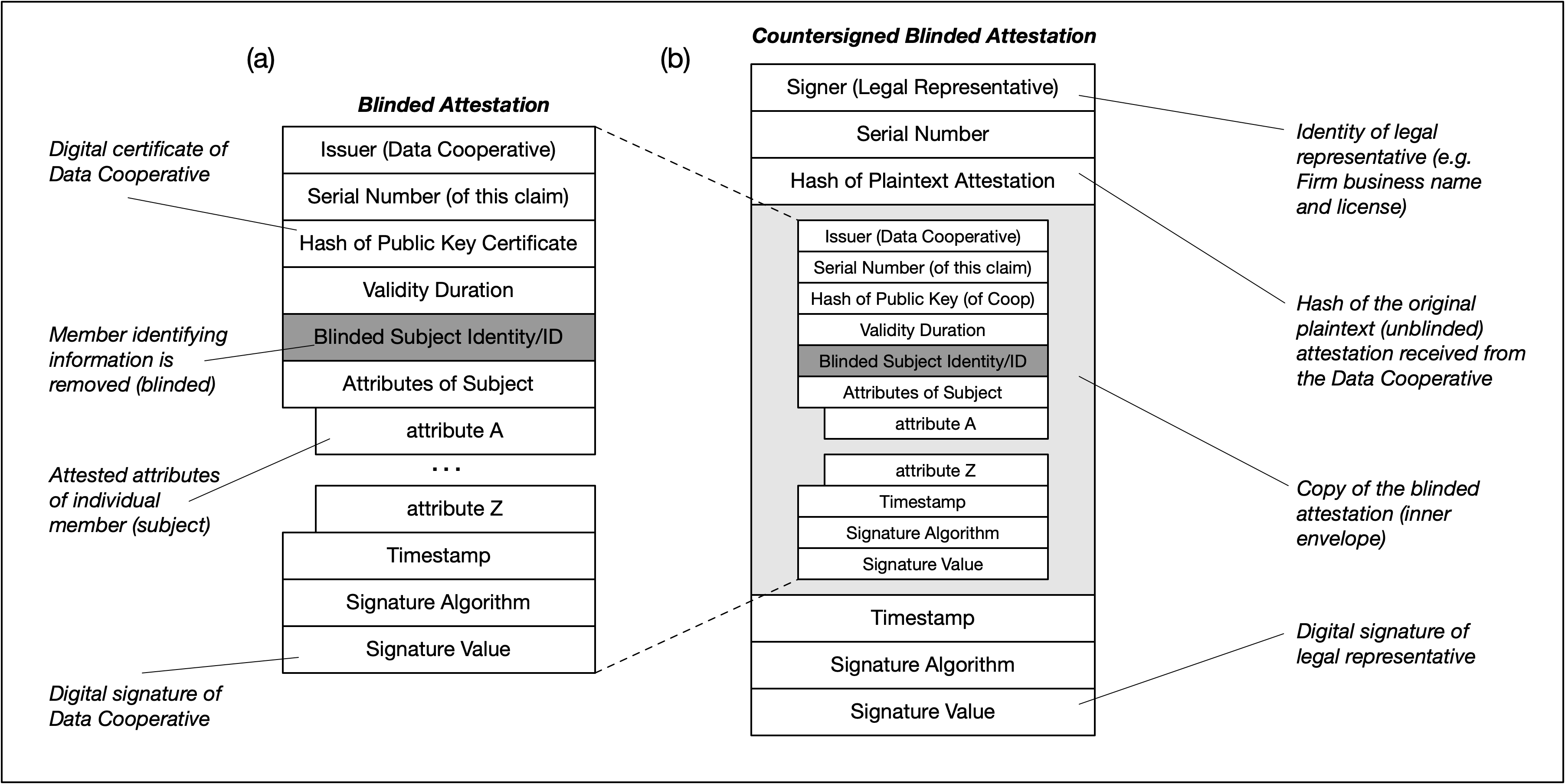}
\caption{Summary of (a) the blinded attestation issued by the Data Cooperative, and (b) the countersigned by the legal representative of the Cooperative (shown using the standard {X.509} and CMS syntax~\cite{RFC5755-X509Attribute-Formatted,RFC5280-formatted,RFC5652-CMS-Formatted}).}
\label{fig:blindedattestation}
\end{figure}

The introduction of the legal representative operating under an attorney-client relationship
with the cooperative provides for the following
(see Figure~\ref{fig:blindedattestation}):
\begin{itemize}

\item	{\em The blinded attestation part (inner part)}: This is the blinded attestation
that hides the identity of the user (member) to whom the attributes pertain.
It is digitally signed by the cooperative,
signifying that it stands 
behind their assertion regarding the truthfulness of the member's attribute.
It is the cooperative who is making this qualitative assertion about its member
and not the legal representative.
This is shown as Figure~\ref{fig:blindedattestation}(a).

\item	{\em The counter signature (outer envelope)}: The counter signature
of the legal representative (e.g. law firm) is performed
on the unmodified copy of the blinded attestation received from the cooperative.

In effect,
it signifies the fact that the legal representative has {\em witnessed} the existence of a matching
pair of (i) a plaintext (unblinded) attestation and (ii) blinded attestation,
both of which were signed by the cooperative.

Secondly,
by countersigning a copy of the blinded attestation the legal representative
indicates that it is acting in its capacity as {\em legal notary}
and that the cooperative has authorized it 
to act as the first point of contact for any legal inquiries regarding the user (member)
whose attribute is stated in the blinded attestation part.
This is shown in (b) in Figure~\ref{fig:blindedattestation}.

Since the legal representative retains (archives) copies of all
three attestations
(the plain attestation signed by the cooperative, 
the blinded attestation signed by the cooperative,
and the countersigned blinded attestation that itself signed)
it can respond to any future queries regarding the identity of the member.

\end{itemize}
It is worthwhile to emphasize that the legal representative (e.g. law firm)
is not attesting to the veracity of the attribute assertion or claim made by the data cooperative.
It is merely witnessing the existence of both the plain attestation and the blinded attestation
(both were signed by the cooperative).


\section{The Travel Rule and Blinded Attestations}
\label{sec:FATF}

The FATF Recommendation~15 of 2018~\cite{FATF-Recommendation15-2018,FATF-Guidance-2019}
defines a {\em virtual asset} as 
a digital representation of value that can be 
digitally traded, or transferred, and can be used for payment or investment purposes. 
Under Recommendation~15
the virtual assets do not include digital representations of fiat currencies, 
securities and other financial assets that are already covered elsewhere in the FATF Recommendations.

The FATF Recommendation~15 also
defines a {\em virtual asset service provider} (VASP) -- most notably the crypto-exchanges -- to be 
a business that conducts one or more of the following activities 
(or  operations for or on behalf of another natural or legal person or business):
(i) exchange between virtual assets and fiat currencies; 
(ii) exchange between one or more forms of virtual assets;
(iii) transfer of virtual assets;
(iv) safekeeping and/or administration of virtual assets or instruments enabling control over virtual assets; and
(v) participation in and provision of financial services related to an issuer's offer and/or sale of a virtual asset.

The implication of the Recommendation~15, among others,
is that crypto-exchanges and other types of VASPs must be able to share the
originator and beneficiary information for virtual asset transactions.
This process -- also known as the {\em Travel Rule} --
originates from under the US Bank Secrecy Act (BSA - 31 USC 5311 - 5330),
which mandates that financial institutions deliver certain types of information
to the next financial institution when a funds transmittal event 
involves more than one financial institution.
This customer information includes (i) originator's name;
(ii) originator's account number (e.g. at the Originator's VASP);
(iii) originator's geographical address, or national identity number, 
or customer identification number (or date and place of birth);
(iv) beneficiary's name;
(v) beneficiary account number (e.g. at the Beneficiary-VASP).

\begin{figure}[t]
\centering
\includegraphics[width=1.0\textwidth, trim={0.0cm 0.0cm 0.0cm 0.0cm}, clip]{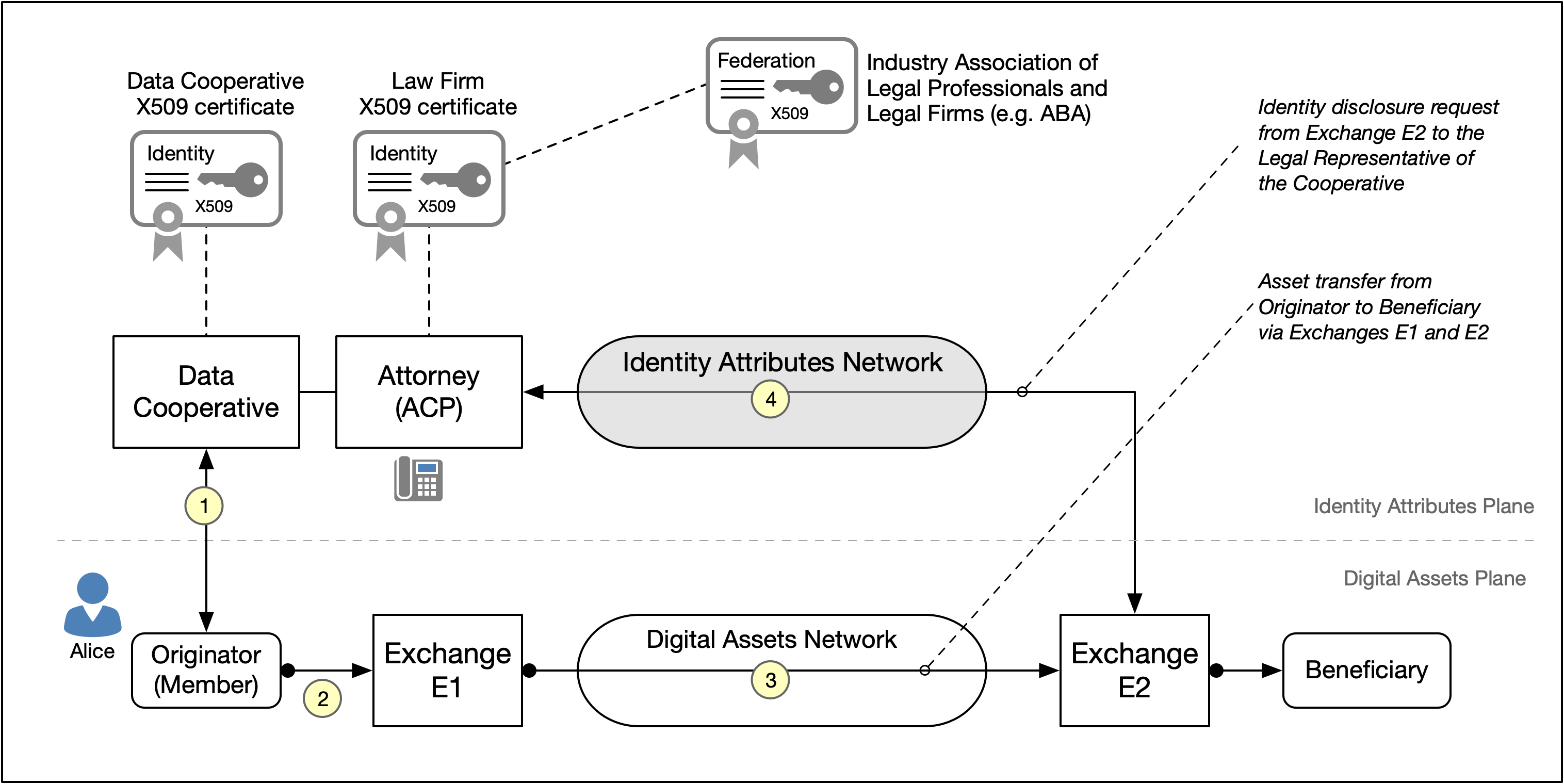}
\caption{Blinded attestations in the context of the Funds Travel Rule and digital assets}
\label{fig:datacoop-vasp}
\end{figure}

We believe the approach of using a legal entity to countersign the blinded attestation
may provide a short-term solution to the Travel Rule problem,
notably the need for the VASPs (crypto-exchanges) to obtain
customer information as required under the BSA regulation.
Figure~\ref{fig:datacoop-vasp} illustrates this use-case scenario,
where the Originator is a member of a data cooperative (Step-1),
and has provided a copy of its blinded attestation
to the exchange that the Originator utilizes (Step-2).

The following is an outline of the use of the blinded attestations
in the context of the Travel Rule (Figure~\ref{fig:datacoop-vasp}):
\begin{itemize}

\item	The Originator must obtain a countersigned blinded attestation
from its data cooperative and its legal representative (Step-1).

\item	The Originator then requests its Exchange E2
to transmit digital assets to the Beneficiary at a different Exchange (Step-2).

\item	When a remote Exchange E2 receives an incoming
asset transfer from an Exchange E1,
it must request a copy of the countersigned blinded attestation (for that Originator)
from the Exchange E1.
This is shown as Step-3 in Figure~\ref{fig:datacoop-vasp}.

The delivery of the countersigned blinded attestation
should be made Out-of-Band,
via a direct secure connection between exchanges E1 and E2 through standardized service interfaces
(see efforts by TRISA to standardize APIs for crypto-exchanges~\cite{TRISA-2019}).

\item	When the Exchange E2 receives the countersigned blinded attestation
it has the option to inquire at two levels (Step-4 in Figure~\ref{fig:datacoop-vasp}).
First,
it can request the legal representative of the cooperative
to validate that the attestation has not been revoked.
Secondly,
it can ask the legal representative to disclose the true identity of the individual member
whose attribute is asserted in the countersigned blinded attestation.
If the Exchange E2 is operating under a legal jurisdiction
compatible with that of the legal representative,
then this second type of request (identity disclosure) may be honored.

\end{itemize}

Readers seeking to obtain further background
on the Travel Rule problem in the context of digital assets 
on decentralized asset networks (i.e. blockchains)
are directed to~\cite{HardjonoLipton2020-FinTech,HardjonoLipton2021-Exchange-NetworksChapter}.


\section{Decentralized Social Networks}
\label{sec:DSNP}

Another potential use-case scenario for blinded attestations is that
of {\em decentralized social networks} (DSN) where the identity privacy of a sender
needs to be protected,
while at the same time
the traffic from unauthenticated parties (e.g. Bots\cite{Kepner2022-ZeroBotnets})
can be filtered out before being transmitted to the social network.

\begin{figure}[t]
\centering
\includegraphics[width=0.9\textwidth, trim={0.0cm 0.0cm 0.0cm 0.0cm}, clip]{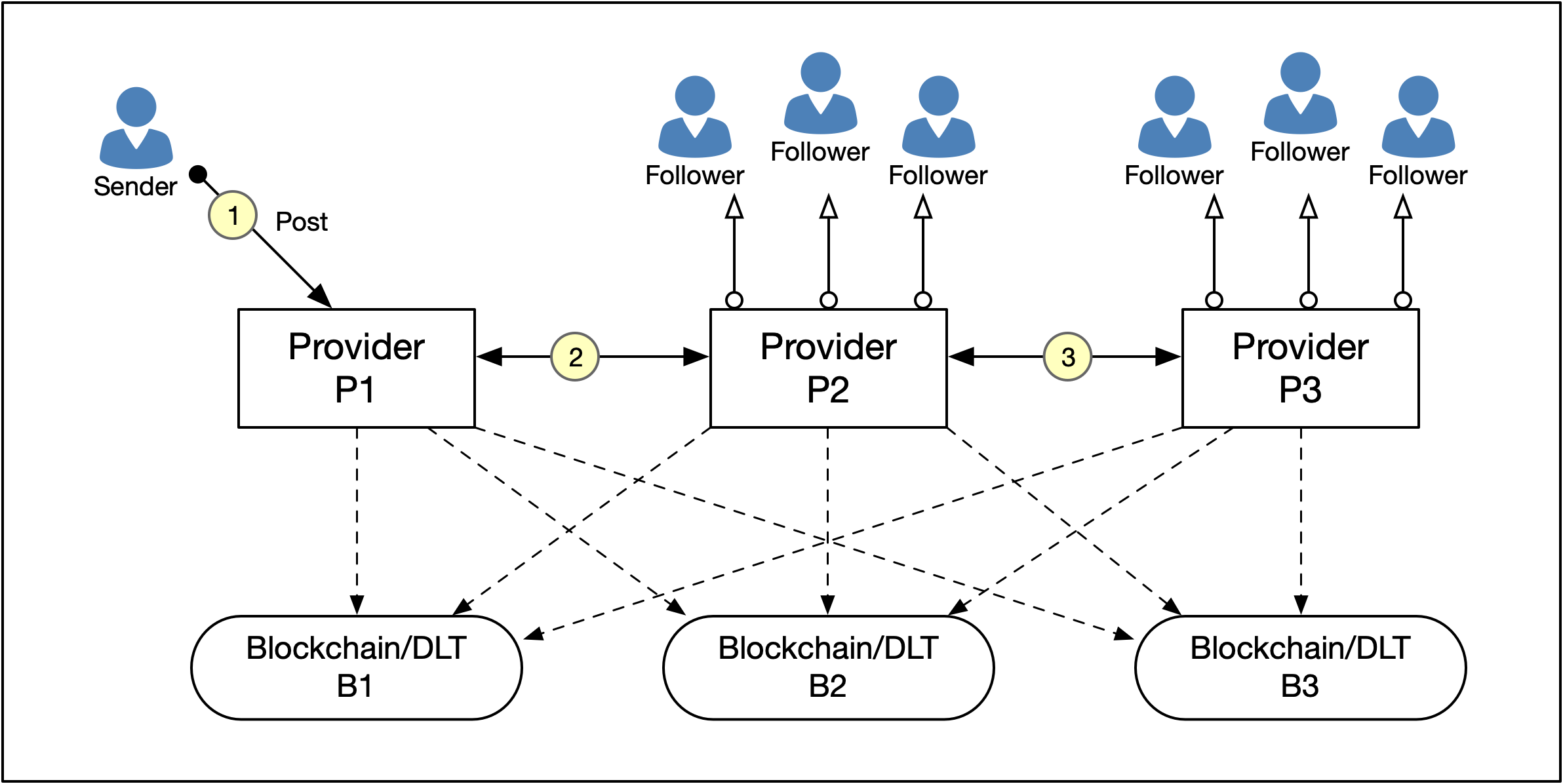}
\caption{Overview of a decentralized social network}
\label{fig:DSN}
\end{figure}

A key concern today regarding many current centralized social networks
is the flooding of the social layer with posts from non-humans (e.g. automated Bots),
often carrying deepfake images or texts that negatively influence the audience.
On the other hand,
many users seek to remain anonymous from parties who are not direct followers
of the user.
That is, a {\em sender} may wish to remain pseudonymous to their {\em followers}
and remain anonymous to non-followers.

Figure~\ref{fig:DSN} summarizes a decentralized model for social networks
where users are free to utilize {\em social media providers}
that forward (receive) posts from a sender to followers attached
through remote providers.
A key goal of these independent providers is to filter out
posts that originate from un-authenticated users or
accounts that appear to be part of a coordinated effort 
(e.g. part of propaganda by state actors~\cite{Erhardt2022-DetectionCoordination}).

However,
one of the main challenges of the emerging decentralized social networks
is the ability for an independent provider
carry-out this task when it has not been the entity who
authenticated the sender.
Thus, for example,
the provider P3 was not the provider
who authenticate the sender (who is attached to provider P1).
However,
the sender has followers who are attached through provider P3
and who seek to (demand to) receive posts from the sender at P1.

We believe there is a role for countersigned blinded attestations
to mitigate some of the risks from bots
while permitting users and providers to be independent
and operate in a decentralized fashion.
The following is an outline of the workflow (see Figure~\ref{fig:DSN}):
\begin{itemize}

\item	A sender who is a member of a data cooperative can request a countersigned blinded attestation
to be issued by the cooperative and its legal representative.
In the case of social networks,
the member may request that Subject Identity/ID field
be substituted with the social media handle name  (e.g.``{\tt @sender}'')  of the member.
(See the grey Subject Identity/ID field in Figure~\ref{fig:blindedattestation}(a)).

\item	Then sender -- who is attached
to provider P1  in Figure~\ref{fig:DSN} and has been authenticated by P1  -- proceeds to request
that P1 store on the blockchain (B1) 
a copy of the sender's countersigned blinded attestation.
This step must be performed before the sender transmits any posts/messages 
through provider P1.

\item	The provider P1 records a copy of the sender's countersigned blinded attestation
to blockchain B1,
using the provider's public-key for that blockchain.
This permits other providers P2 and P3 to validate that the
the entry/record on blockchain B1 was made by provider P1.
The countersigned blinded attestation is therefore publicly readable
on blockchain B1.

\item	Each time the sender transmits a post/message via the provider P1 (Line-1 in Figure~\ref{fig:DSN}),
the provider P1 will compute a hash of the post
and record this hash on blockchain B1 
together with a pointer to the record/block on B1 storing
sender's countersigned blinded attestation.
After this step,
the provider P1 propagates the post/message to providers P2 and P3
where some followers are attached (Line-2 and Line-3 in Figure~\ref{fig:DSN}).

\item	When a remote provider (P2 or P3) seeks to evaluate
an incoming post/message forwarded by provider P1,
the providers P2 and P3 must search on blockchain B1
for a matching hash of the message.
That is,
the providers P2 and P3 must re-compute the hash of the received post/message
from P1 and use this to search blockchain B1.

\item	Using the matching hash found on blockchain B1
the providers (P2 and P3) are then able to locate
the countersigned blinded attestation of the sender on blockchain B1.
These remote providers P2 and P3 may optionally
port a copy of the countersigned blinded attestation onto their own blockchains B2 and B3 respectively.

\item	In cases where the providers P1, P2 or P3 requests
the disclosure of the sender's identity,
the providers must contact the legal representative of the sender
(see previous Figure~\ref{fig:datacoop-vasp} at Step-4).
The identity of the legal representative is stated 
in the plaintext portion of the countersigned blinded attestation
(top part of Figure~\ref{fig:blindedattestation}(b)).

\end{itemize}

What is obtained from the above is the following:
\begin{enumerate}
\item	{\em Mutual reliance among decentralized providers}: 
A non-origin provider (i.e. P2 and P3) is willing to receive a sender's forwarded posts because
the origin provider (i.e. P1) has authenticated the sender and has 
recorded identity-related metadata about the sender
(i.e. the sender's countersigned blinded attestation) on its public blockchain B1.

This need to establish mutual reliance
is reminiscent of the {\em source address validation} (SAV) problem
among ISPs in IP routing.

\item	{\em Enhanced filter capabilities to address the bots problem}:
Each provider has an enhanced filtering capability
because it has an additional parameter upon which
to base its filter policies.
This parameter is based on the existence
of the countersigned blinded attestations
on the blockchains of the providers.

\item	{\em Privacy for users and method of recourse for providers}:
The sender's identity remains hidden until such time disclosure is requested
via the cooperative's legal representative.
A provider has a point of contact for the sender
in the case that disputes occur.

\end{enumerate}
Other features can be added to the attributes list within the blinded attestations
(Figure~\ref{fig:blindedattestation}(a)),
such as a {\em recovery public-key} in the case where the sender's account at provider P1
has been lost (e.g. hacked or stolen).
The recovery key-pair is to be used by the sender to notify all the providers,
triggering the origin provider P1 to record
a fresh countersigned blinded attestations onto blockchain B1.

There are currently a number of 
projects on decentralized social networks (e.g. Project Liberty DSNP~\cite{DSNP-Whitepaper-2020})
that could benefit from the blinded attestations approach outlined above.


\section{Conclusion}
\label{sec:Conclusion}

In the current work we extend the fiduciary relationship between a data cooperative
and its members to include the cooperative being an issuer of attribute attestations 
for its members.
An individual member can therefore request the cooperative 
to issue signed attestations in a digital format that can later
be utilized by the individual to obtain services at third parties
who need to perform a risk assessment about that individual.

Since one of the main propositions of the cooperative model
is to protect the data privacy of members,
we proposed the use of {\em blinded attestations}
in which the identity of the member (data subject) is removed 
from any attestations issued by the cooperative.
This blinded attestation is then countersigned by a legal representative (e.g. law firm)
of the cooperative,
making use of the attorney-client relationship with the cooperative
This enables the legal representative to
henceforth become the legal point of contact
for inquiries regarding the individual related to the attribute being attested.

The blinded attestation approach has applicability in the context of the Travel Rule
for the transfer of digital assets under the FATF Recommendation~15 and other related regulations
for funds transfer (e.g. BSA~1996).
Additionally,
the same blinded attestation approach can be utilized to protect user privacy
in the emerging decentralized social networks.

~~\\


\section*{Acknowledgements}
\label{sec:Acknowledgement}

We thank Robert Mahari from MIT Human Dynamics for inputs and reviews.

\small


\begin{thebibliography}{10}
\providecommand{\url}[1]{#1}
\csname url@samestyle\endcsname
\providecommand{\newblock}{\relax}
\providecommand{\bibinfo}[2]{#2}
\providecommand{\BIBentrySTDinterwordspacing}{\spaceskip=0pt\relax}
\providecommand{\BIBentryALTinterwordstretchfactor}{4}
\providecommand{\BIBentryALTinterwordspacing}{\spaceskip=\fontdimen2\font plus
\BIBentryALTinterwordstretchfactor\fontdimen3\font minus
  \fontdimen4\font\relax}
\providecommand{\BIBforeignlanguage}[2]{{%
\expandafter\ifx\csname l@#1\endcsname\relax
\typeout{** WARNING: IEEEtran.bst: No hyphenation pattern has been}%
\typeout{** loaded for the language `#1'. Using the pattern for}%
\typeout{** the default language instead.}%
\else
\language=\csname l@#1\endcsname
\fi
#2}}
\providecommand{\BIBdecl}{\relax}
\BIBdecl

\bibitem{PentlandHardjono2020-DataCooperativesChapter}
A.~Pentland and T.~Hardjono, ``{Building Data Cooperatives},'' in
  \emph{Building the New Economy: Data as Capital}, A.~Pentland, A.~Lipton, and
  T.~Hardjono, Eds.\hskip 1em plus 0.5em minus 0.4em\relax MIT Press, 2021, pp.
  19--33.

\bibitem{HardjonoPentland2019-DataCooperatives-Arxiv}
\BIBentryALTinterwordspacing
------, ``{Data Cooperatives: Towards a Foundation for Decentralized Personal
  Data Management},'' May 2019. [Online]. Available:
  \url{https://arxiv.org/abs/1905.08819}
\BIBentrySTDinterwordspacing

\bibitem{openPDS2014PLOS}
Y.~A. {de~Montjoye}, E.~Shmueli, S.~Wang, and A.~Pentland, ``{openPDS}:
  {P}rotecting the {P}rivacy of {M}etadata through {SafeAnswers},'' \emph{PLoS
  ONE 9(7)}, pp. 13--18, July 2014,
  https://doi.org/10.1371/journal.pone.0098790.

\bibitem{Balkin2016}
J.~M. Balkin, ``{I}nformation {F}iduciaries and the {F}irst {A}mendment,''
  \emph{UC Davis Law Review}, vol.~49, no.~4, pp. 1183--1234, April 2016.

\bibitem{Madden2014}
M.~Madden, ``{P}ublic {P}erceptions of {P}rivacy and {S}ecurity in the
  {P}ost-{S}nowden {E}ra,'' November 2014,
  http://www.pewinternet.org/2014/11/12/public-privacy-perceptions/.

\bibitem{Webb2015}
A.~Webb, ``{T}he {T}ech {T}rends {Y}ou {Can't} {I}gnore in {2015},'' January
  2015, https://hbr.org/2015/01/the-tech-trends-you-cant-ignore-in-2015.

\bibitem{Maler2015a}
E.~Maler, ``{E}xtending the {P}ower of {C}onsent with {U}ser-{M}anaged
  {A}ccess: {A} {S}tandard {A}rchitecture for {A}synchronous, {C}entralizable,
  {I}nternet-{S}calable {C}onsent,'' in \emph{Proc. 2015 IEEE Security and
  Privacy Workshops}, San Jose, May 2015, dOI: 10.1109/SPW.2015.34.

\bibitem{Anthem2015}
\BIBentryALTinterwordspacing
R.~Abelson and M.~Goldstein, ``Millions of {A}nthem customers targeted in
  cyberattack,'' \emph{New York Times}, February 2015. [Online]. Available:
  \url{https://www.nytimes.com/2015/02/05/business/hackers-breached-data-of-millions-insurer-says.html}
\BIBentrySTDinterwordspacing

\bibitem{Equifax2017}
\BIBentryALTinterwordspacing
T.~S. Bernard, T.~Hsu, N.~Perlroth, and R.~Lieber, ``Equifax says cyberattack
  may have affected 143 million in the {U.S.}'' \emph{New York Times},
  September 2017. [Online]. Available:
  \url{https://www.nytimes.com/2017/09/07/business/equifax-cyberattack.html}
\BIBentrySTDinterwordspacing

\bibitem{WEF2011}
{World Economic Forum}, ``{P}ersonal {D}ata: {T}he {E}mergence of a {N}ew
  {A}sset {C}lass,'' 2011,
  http://www.weforum.org/reports/personal-data-emergence-new-asset-class.

\bibitem{GDPR}
{European Commission}, ``Regulation {(EU)} 2016/679 of the {E}uropean
  {P}arliament and of the {C}ouncil of 27 {A}pril 2016 on the protection of
  natural persons with regard to the processing of personal data and on the
  free movement of such data ({G}eneral {D}ata {P}rotection {R}egulation),''
  \emph{Official Journal of the European Union}, vol. L119, pp. 1--88, 2016.

\bibitem{OAIC-Australia-2018}
\BIBentryALTinterwordspacing
{OAIC}, ``{Australian entities and the EU General Data Protection Regulation
  (GDPR)},'' Australian Government Office of the Australian Information
  Commissioner, Tech. Rep., 2018, accessed 23 August 2021. [Online]. Available:
  \url{https://www.oaic.gov.au/privacy/guidance-and-advice/australian-entities-and-the-eu-general-dataprotection-regulation/}
\BIBentrySTDinterwordspacing

\bibitem{MiCA2023}
{European Commission}, ``Regulation {(EU)} 22023/1114 of the {E}uropean
  {P}arliament and of the {C}ouncil of 27 {A}pril 2016 on {markets in
  crypto-assets, and amending Regulations (EU) No 1093/2010 and (EU) No
  1095/2010 and Directives 2013/36/EU and (EU) 2019/1937},'' \emph{Official
  Journal of the European Union}, vol. L150, pp. 1--166, June 2023.

\bibitem{Pentland2015-short}
A.~Pentland, \emph{{ Social Physics}}.\hskip 1em plus 0.5em minus 0.4em\relax
  Penguin Books, 2015.

\bibitem{PentlandHardjono2018a}
A.~Pentland and T.~Hardjono, ``Digital identity is broken: Here is a way to fix
  it,'' \emph{Wall Street Journal}, April 2018,
  https://www.wsj.com/articles/digital-identity-is-broken-heres-a-way-to-fix-it-1522782822.

\bibitem{RFC1991-formatted}
\BIBentryALTinterwordspacing
D.~Atkins, W.~Stallings, and P.~Zimmermann, ``{PGP} {M}essage {E}xchange
  {F}ormats,'' August 1996, {IETF}~{S}tandard~{RFC1991}. [Online]. Available:
  \url{http://tools.ietf.org/rfc/rfc1991.txt}
\BIBentrySTDinterwordspacing

\bibitem{PGP-Key-Signing-IETF-1999}
\BIBentryALTinterwordspacing
T.~Tso, ``{IETF PGP Key Signing announcement (IETF Minneapolis March 1999)},''
  March 1999. [Online]. Available:
  \url{https://mailarchive.ietf.org/arch/msg/ietf/SDcqTgHwOpmooLxxLzItLMxzwNA/}
\BIBentrySTDinterwordspacing

\bibitem{SAMLcore}
{OASIS}, ``{A}ssertions and {P}rotocols for the {OASIS} {S}ecurity {A}ssertion
  {M}arkup {L}anguage ({SAML}) {V2.0},'' March 2005, available on
  {http://docs.oasisopen.org/security/ saml/v2.0/ saml-core-2.0-os.pdf}.

\bibitem{Sporny2022}
\BIBentryALTinterwordspacing
M.~Sporny, D.~Longley, and D.~Chadwick, ``{V}erifiable {C}redentials {D}ata
  {M}odel {1.1},'' {W3C}, {W3C} {R}ecommendation, March 2022. [Online].
  Available: \url{https://www.w3.org/TR/vc-data-model/}
\BIBentrySTDinterwordspacing

\bibitem{OIDC1.0}
N.~Sakimura, J.~Bradley, M.~Jones, B.~de~Medeiros, and C.~Mortimore, ``{OpenID}
  {C}onnect {C}ore {1.0},'' OpenID Foundation, Technical Specification {v1.0}
  -- Errata Set 1, November 2014,
  http://openid.net/specs/openid-connect-core-1\_0.html.

\bibitem{SAMLglossary}
{OASIS}, ``{G}lossary for the {OASIS} {S}ecurity {A}ssertion {M}arkup
  {L}anguage ({SAML}) {V2.0},'' March 2005, available on
  {http://docs.oasis-open.org/security/ saml/v2.0/samlglossary- 2.0-os.pdf}.

\bibitem{ABA2012}
{American Bar Association}, ``{A}n {O}verview of {I}dentity {M}anagement:
  {S}ubmission for {UNCITRAL} {C}ommission {45th} {S}ession,'' {ABA Identity
  Management Legal Task Force}, May 2012, available on
  http://meetings.abanet.org/ webupload/commupload/ CL320041/relatedresources/
  ABA-Submission-to-UNCITRAL.pdf.

\bibitem{FATF-Recommendation15-2018}
{FATF}, ``{I}nternational {S}tandards on {C}ombating {M}oney {L}aundering and
  the {F}inancing of {T}errorism and {P}roliferation,'' Financial Action Task
  Force (FATF), {FATF}~{R}evision of {R}ecommendation~{15}, October 2018,
  available at:
  http://www.fatf-gafi.org/publications/fatfrecommendations/documents/fatf-recommendations.html.

\bibitem{Chaum81}
D.~L. Chaum, ``Untraceable electronic mail, return addresses, and digital
  pseudonyms,'' \emph{Communications of the ACM}, vol.~24, no.~2, pp. 84--88,
  February 1981.

\bibitem{Brands93a}
S.~Brands, ``Untraceable off-line cash in wallets with observers,'' in
  \emph{{CRYPTO'93} Proceedings of the 13th Annual International
  Cryptology}.\hskip 1em plus 0.5em minus 0.4em\relax Springer-Verlag, 1993,
  pp. 302--318.

\bibitem{Camenisch2002}
J.~Camenisch and E.~{Van~Herreweghen}, ``Design and implementation of the
  {Idemix} anonymous credential system,'' in \emph{Proceedings of the 9th ACM
  conference on Computer and communications security}.\hskip 1em plus 0.5em
  minus 0.4em\relax ACM, 2002, pp. 21--30.

\bibitem{BrickellLi2012}
E.~Brickell and J.~Li, ``{E}nhanced {P}rivacy {ID}: a {D}irect {A}nonymous
  {A}ttestation {S}cheme with {E}nhanced {R}evocation {C}apabilities,''
  \emph{IEEE Transactions on Dependable and Secure Computing}, vol.~9, no.~3,
  pp. 345--360, 2012.

\bibitem{ABA1983-Attroney-Client}
\BIBentryALTinterwordspacing
{American Bar Association}, ``{Rules of Professional Conduct Rule 1.6:
  Confidentiality of information},'' 1983. [Online]. Available:
  \url{https://www.americanbar.org/groups/professional_responsibility/publications/model_rules_of_professional_conduct/}
\BIBentrySTDinterwordspacing

\bibitem{Chaum:1990:UEC:88314.88969}
\BIBentryALTinterwordspacing
D.~Chaum, A.~Fiat, and M.~Naor, ``Untraceable electronic cash,'' in
  \emph{Proceedings on Advances in Cryptology}, ser. CRYPTO '88.\hskip 1em plus
  0.5em minus 0.4em\relax New York, NY, USA: Springer-Verlag New York, Inc.,
  1990, pp. 319--327. [Online]. Available:
  \url{http://dl.acm.org/citation.cfm?id=88314.88969}
\BIBentrySTDinterwordspacing

\bibitem{NIST-DSS-2023}
\BIBentryALTinterwordspacing
{NIST}, ``{D}igital {S}ignature {S}tandard ({DSS}),'' National Institute of
  Standards and Technology (NIST), {NIST}~{FIPS~186-5}, February 2023.
  [Online]. Available: \url{https://doi.org/10.6028/NIST.FIPS.186-5}
\BIBentrySTDinterwordspacing

\bibitem{RFC5652-CMS-Formatted}
\BIBentryALTinterwordspacing
R.~Housley, ``{Cryptographic Message Syntax (CMS)},'' September 2009,
  {IETF}~{S}tandard~{RFC5652}. [Online]. Available:
  \url{https://datatracker.ietf.org/doc/html/rfc5652}
\BIBentrySTDinterwordspacing

\bibitem{RFC7517-JWK-Formatted}
\BIBentryALTinterwordspacing
M.~Jones, ``{JSON Web Key (JWK)},'' May 2015, {IETF}~{S}tandard~{RFC7517}.
  [Online]. Available: \url{https://www.rfc-editor.org/rfc/rfc7517}
\BIBentrySTDinterwordspacing

\bibitem{RFC7515-JOSE-Formatted}
\BIBentryALTinterwordspacing
M.~Jones, J.~Bradley, and N.~Sakimura, ``{JSON Web Signature (JWS)},'' May
  2017, {IETF}~{S}tandard~{RFC7515}. [Online]. Available:
  \url{https://tools.ietf.org/html/rfc7515}
\BIBentrySTDinterwordspacing

\bibitem{RFC5755-X509Attribute-Formatted}
\BIBentryALTinterwordspacing
S.~Farrell, R.~Housley, and S.~Turner, ``{An Internet Attribute Certificate
  Profile for Authorization},'' January 2010, {IETF}~{S}tandard~{RFC5755}.
  [Online]. Available: \url{https://datatracker.ietf.org/doc/html/rfc5755}
\BIBentrySTDinterwordspacing

\bibitem{RFC5280-formatted}
\BIBentryALTinterwordspacing
D.~Cooper, S.~Santesson, S.~Farrell, S.~Boeyen, R.~Housley, and W.~Polk,
  ``{I}nternet {X.509} {P}ublic {K}ey {I}nfrastructure {C}ertificate and
  {C}ertificate {R}evocation {L}ist ({CRL}) {P}rofile,'' May 2008,
  {IETF}~{S}tandard~{RFC5280}. [Online]. Available:
  \url{http://tools.ietf.org/rfc/rfc5280.txt}
\BIBentrySTDinterwordspacing

\bibitem{FATF-Guidance-2019}
{FATF}, ``{G}uidance for a {R}isk-{B}ased {A}pproach to {V}irtual {A}ssets and
  {V}irtual {A}sset {S}ervice {P}roviders,'' Financial Action Task Force
  (FATF), {FATF}~{G}uidance, June 2019, available at:
  www.fatf-gafi.org/publications/fatfrecommendations/documents/Guidance-RBA-virtual-assets.html.

\bibitem{TRISA-2019}
\BIBentryALTinterwordspacing
{TRISA}, ``{T}ravel {R}ule {I}nformation {S}haring {A}rchitecture for {V}irtual
  {A}sset {S}ervice {P}roviders {(TRISA)} -- {V}ersion~5,'' December 2019.
  [Online]. Available:
  \url{https://trisacrypto.github.io/white-papers/white-paper-trisa-v5.pdf}
\BIBentrySTDinterwordspacing

\bibitem{HardjonoLipton2020-FinTech}
\BIBentryALTinterwordspacing
T.~Hardjono, A.~Lipton, and A.~Pentland, ``{T}owards a {P}ublic {K}ey
  {M}anagement {F}ramework for {V}irtual {A}ssets and {V}irtual {A}sset
  {S}ervice {P}roviders,'' \emph{{J}ournal of {FinTech}}, vol.~1, no.~1, 2020,
  available at {https://arxiv.org/pdf/1909.08607}. [Online]. Available:
  \url{https://doi.org/10.1142/S2705109920500017}
\BIBentrySTDinterwordspacing

\bibitem{HardjonoLipton2021-Exchange-NetworksChapter}
------, ``{Exchange Networks for Virtual Assets},'' in \emph{Building the New
  Economy: Data as Capital}, A.~Pentland, A.~Lipton, and T.~Hardjono,
  Eds.\hskip 1em plus 0.5em minus 0.4em\relax MIT Press, 2021.

\bibitem{Kepner2022-ZeroBotnets}
\BIBentryALTinterwordspacing
J.~Kepner, J.~Bernays, S.~Buckley, K.~Cho, C.~Conrad, L.~Daigle, K.~Erhardt,
  V.~Gadepally, B.~Greene, M.~Jones, R.~Knake, B.~M. Maggs, P.~Michaleas, C.~R.
  Meiners, A.~Morris, A.~Pentland, S.~Pisharody, S.~Powazek, A.~Prout,
  P.~Reiner, K.~Suzuki, K.~Takahashi, T.~Tauber, L.~Walker, and D.~Stetson,
  ``{Zero Botnets: An Observe-Pursue-Counter Approach}.'' \emph{CoRR}, vol.
  abs/2201.06068, 2022. [Online]. Available:
  \url{https://arxiv.org/abs/2201.06068}
\BIBentrySTDinterwordspacing

\bibitem{Erhardt2022-DetectionCoordination}
\BIBentryALTinterwordspacing
K.~Erhardt and A.~Pentland, ``{Detection Of Coordination Between State-Linked
  Actors},'' in \emph{Social, Cultural, and Behavioral Modeling: 15th
  International Conference, {SBP-BRiMS 2022}}.\hskip 1em plus 0.5em minus
  0.4em\relax Springer-Verlag, 2022, p. 144–154. [Online]. Available:
  \url{https://doi.org/10.1007/978-3-031-17114-7_14}
\BIBentrySTDinterwordspacing

\bibitem{DSNP-Whitepaper-2020}
\BIBentryALTinterwordspacing
{Project~Liberty}, ``{ Decentralized Social Networking Protocol },'' October
  2022. [Online]. Available:
  \url{https://github.com/LibertyDSNP/papers/blob/main/whitepaper/dsnp_whitepaper.pdf}
\BIBentrySTDinterwordspacing

\end{thebibliography}


\end{document}